\documentclass{iau} 
\usepackage{graphicx}
\usepackage{floatrow}

\title[Stellar ages] 
{The ages of (the oldest) stars}

\author[M. Catelan]   
{M. Catelan$^{1,2}$} 

\affiliation{$^1$Pontificia Universidad Cat\'olica de Chile, Facultad de F\'{i}sica, Instituto de Astrof\'\i sica,\\ Av. Vicu\~{n}a Mackenna 4860, 
       782-0436 Macul, Santiago, Chile\\ email: {\tt mcatelan@astro.puc.cl}\\[\affilskip]
$^2$Millennium Institute of Astrophysics, Santiago, Chile}

\pubyear{2017}
\volume{334}  
\jname{Rediscovering our Galaxy}
\editors{C. Chiappini, I. Minchev, E. Starkenburg, \\ \& M. Valentini, eds.}
\begin{document}

\maketitle

\begin{abstract}
Much progress has been achieved in the age-dating of old stellar systems, and even of individual stars in the field, in the more than sixty years since the evolution of low-mass stars was first correctly described. In this paper, I provide an overview of some of the main methods that have been used in this context, and discuss some of the issues that still affect the determination of accurate ages for the oldest stars.


\end{abstract}

\firstsection 

\section{Introduction}\label{sec:intro}

Along with their masses, the ages of stars constitute one of their most important, if elusive, parameters. For most stars, age is a quantity that can only be measured through indirect means, using stellar evolution models to ``read'' the stars' internal ``clocks.'' This is typically achieved by comparing model predictions with observations of the stars' surface properties~-- be it luminosity/gravity and temperature/color, in most traditional methods, or the power spectrum of oscillations, in the case of asteroseismology. The idea behind this approach is that, as a star progressively burns its nuclear fuel as it ages, its surface properties will change as well. Stellar ages thus depend crucially on the reliability of our models to properly describe the physical conditions in stellar interiors. 

Both A. Noels and A. Weiss, in these same proceedings, address some of the current uncertainties affecting the input physics that is used to compute such models, and their impact on the inferred ages. Accordingly, I will limit myself to providing an overview of some of the techniques that have been used to age-date stars, focusing on the oldest ones, and especially those in globular star clusters. For a broader overview of the age dating of individual stars covering a wide range in ages, the reader is referred to \cite[Soderblom (2010)]{ds10}. Since, as pointed out by \cite[Soderblom]{ds10}, our only available {\em fundamental} calibrator is the Sun, let me start by reviewing the latest available information regarding its age.

\section{The Sun, our sole calibrator}\label{sec:sun}

For our own star, unlike any other, an age estimate can be obtained by borrowing the techniques of {\em geochemistry}. By dating the oldest rocks that are directly accessible to us, as in the case of meteorites~-- fossil remnants of the earliest stages of our Solar System's formation~-- we are able to obtain a detailed chronology of the formation of the Solar System (e.g., \cite[Chaussidon 2007]{mc07}; \cite[Amelin \& Ireland 2013]{ai13}; \cite[Connelly \etal\ 2017]{jcea17}; and references therein). This dating process is carried out using a variety of radioactive isotopes. The long-lived ones, with half-lives of order several Gyr (e.g., $^{87}$Rb, $^{187}$Re, $^{235}$U, $^{238}$U, $^{232}$Th), are the most suitable. The relative proportions between these ``parent species'' and their ultimate decay products (the ``daughter species'') can then be used as a ``clock'' to estimate the age (e.g., \cite[Pagel 2009]{bp09}, and references therein). Very recently, using the Th-U ``clock,'' \cite[Connelly \etal\ (2017)]{jcea17} found an age of $4.56730 \pm 0.00016$~Gyr for the Solar System; the tiny error bar is typical in such studies. However, the authors do caution that the accuracy of these results ``within the stated 0.16~Myr uncertainty'' depends on the calibration of the available tracers and treatment of fractionation for the several different Pb (daughter) isotopes. Note, in addition, that an age difference of a few Myr may exist between the moment the Sun's latest progenitor supernova (or other source of $r$-process elements) ejected its radioactive products into the interstellar medium and the  incorporation of these products into the proto-Solar System nebula. Another Myr or so may also go by, between the formation of a pre-stellar dense core in the parent cloud and the arrival of a Sun-like star at the pre-main sequence (MS) ``birthline'' (e.g., \cite[Wuchterl \& Klessen 2001]{wk01}; \cite[Reipurth 2005]{br05}; \cite[Dauphas \& Chaussidon 2011]{dc11}). Finally, another $\sim 50$~Myr may elapse, before the star settles on the MS (e.g., \cite[Iben 1965]{ii65}; \cite[Sackmann et al. 1993]{ijsea93}). Thus, the exact value of the ``solar age'' may differ somewhat from the quoted value, depending on which point in time one chooses as reference. In addition to that, it should be noted that the isotopes that we use in these measurements do not necessarily owe their origin to a single supernova (or other source of $r$-process elements) event that took place immediately before/during the earliest stages of the Solar System's formation, but rather may have been more continuously supplied by the process of chemical evolution that has been enriching the gas since the earliest stages in the life of the Milky Way \cite[(e.g., Clayton 1988)]{dc88}. The quoted error bar is thus not fully representative of the different sources of uncertainty that affect the determination of the solar age.

\section{Globular cluster ages}\label{sec:gc}

The {\em MS turnoff (TO) point}, which closely traces the point in a star's life at which hydrogen is exhausted in the core, is the primary ``clock'' that is used in dating star  clusters. Different methods variously use the TO luminosity or color as main indicator; one often refers to those as ``vertical'' or ``horizontal'' methods, respectively. The stellar mass at the TO point is also predicted to be a sensitive function of age, and so, when the mass can be directly measured, using detached eclipsing binaries, an estimate of the cluster's age can also be obtained, independent of the uncertainties affecting color transformations and bolometric corrections. Reviews of the subject can be found in \cite[VandenBerg \etal\ (1996)]{vbs96} and \cite[Sarajedini \etal\ (1997)]{asea97}, among many others. A nice account of the early history of globular cluster (GC) age dating is given in \cite[Demarque \etal\ (1991)]{pdea91}. 

The so-called {\underline{\em ``vertical methods''} work by registering specific evolutionary sequences in a color-magnitude diagram (CMD) to reference, standard loci. In this way, the CMD is ``anchored'' into the absolute domain, the {\em distance modulus} obtained, and the absolute magnitude of the TO point~-- hence the age~-- can be directly read. Sometimes, rather than the TO proper (or solely), other nearby positions in the CMD are used (e.g., \cite[Chaboyer \etal\ 1996b]{bcea96b}; \cite[Buonanno \etal\ 1998]{rbea98}; \cite[VandenBerg \etal\ 2013]{dvea13}). If the chemical composition is also known, one is then able to obtain the cluster age. 

In the so-called {\em $\Delta V$ method}, the reference locus is the horizontal branch (HB; \cite[Iben 1974]{ii74}; \cite[Sandage \etal\ 1981]{asea81}). Numerous calibrations, both empirical and theoretical, are available for the dependence of the $V$-band absolute magnitude of the HB with metallicity. For RR Lyrae-rich clusters, the period-luminosity-metallicity (PLZ) relation can also be used, though only for passbands redder than $R$ (\cite[Catelan \etal\ 2004]{mcea04}; \cite[Neeley \etal\ 2017]{jnea17}). 

Many different techniques have been proposed to obtain the absolute magnitudes of HB stars, and RR Lyrae stars in particular; a review can be found in \cite[Cacciari \& Clementini (2003)]{cc03}. The most direct such method is, naturally, the use of trigonometric parallaxes. RR Lyrae stars are, however, typically quite distant, and this has, until recently, limited the sample with reliable parallax measurements to basically one star, namely RR Lyr itself (\cite[e.g., Catelan 2009a, and references therein]{mc09}). With the Gaia Data Release 1 (DR1; \cite[Brown \etal\ 2016]{abea16}), parallaxes have already become available for as many as 200 RR Lyrae stars. This allowed \cite[Clementini \etal\ (2017)]{gcea17} to revisit the zero point (but not the slope) of these relations. However, the different technical procedures adopted by these authors to obtain the zero point of the relation lead to different results. For instance, for the $M_V({\rm RRL})-{\rm [Fe/H]}$ relation, the zero points range from $0.82 \pm 0.04$ to $1.01 \pm 0.04$. Unfortunately, this implies an uncertainty in TO ages of order 2~Gyr. In addition to more firmly establishing the zero point of such relations, it will be important to revisit the metallicity dependence using data from future Gaia releases, considering its direct impact upon the age-metallicity relation, as derived using the $\Delta V$ method. In this sense, the direct determination of GC distances using Gaia will also provide a ``shortcut'' to estimates of their TO absolute magnitudes, hence ages (\cite[Pancino \etal\ 2017]{epea17}). 

One note of caution, regarding the application of $M_V({\rm HB})-{\rm [Fe/H]}$ calibrations based on field stars to GC work, is in order. As well known, GCs have been found to harbor multiple populations, with so-called ``second-generation stars'' differing from the first-generation ones by showing evidence of additional proton-capture nucleosynthesis, in the form of Na-O and Mg-Al anticorrelations (\cite[see Gratton et al. 2012, for a review]{rgea12}). Among field stars, on the other hand, the presence of such second-generation stars is small at best. Thus, the possibility exists that RR Lyrae stars in some GCs, but not in the field, may be enhanced in helium, and thus be overluminous in comparison with the latter. In fact, \cite[VandenBerg \etal\ (2016)]{dvea16} recently favored a small amount of helium enhancement for the RR Lyrae stars in M15 (NGC~7078), but not in M3 (NGC~5272) or M92 (NGC~6341). Taking such helium abundance variations into account, they obtain a similar age for all three clusters, $\approx 12.5$~Gyr, depending on the assumed CNO abundances. 

While using HB stars is particularly widespread, this is by no means the only way to ``anchor'' a CMD. In this sense, the {\em white dwarf (WD) cooling sequence method} uses the cooling sequence of WDs in the CMD to infer the cluster distances, hence their TO ages. The method is calibrated using samples of nearby WDs with known trigonometric parallaxes. The cluster photometry is usually obtained using the {\em Hubble Space Telescope} ({\em HST}), and typically restricted to the nearest objects, due to the intrinsic faintness of WDs. Applying the method to the GC NGC~6752, \cite[Renzini et al. (1996)]{area96} advocated a cluster age of $15 \pm 1.5 \pm 0.5$~Gyr (statistical and systematic uncertainties indicated), while \cite[Zoccali et al. (2001)]{mzea01}, using the same method, found an age of $13 \pm 2.5$~Gyr for 47~Tuc (NGC~104). It is important to note that the {\underline{\em WD luminosity function}} is also a powerful indicator of cluster age, and one that has been frequently used as well~-- though again, only for the nearest objects. \cite[Hansen \etal\ (2013)]{bhea13} obtained an age of $9.9 \pm 0.7$~Gyr for 47~Tuc using this method, whereas \cite[Campos \etal\ (2016)]{fcea16} favor instead $11.31^{+0.36}_{-0.17}$~Gyr; both studies agree, however, that NGC~6397 is about 2~Gyr older than 47~Tuc, implying a significant age-metallicity relation for Galactic globulars. A number of age-sensitive diagnostics based on WD stars that can be exploited in the near-infrared have been discussed by \cite[Bono \etal\ (2013)]{gbea13}. For a discussion of some of the uncertainties affecting such WD-based methods, we refer the reader to \cite[Salaris (2013)]{ms13}. A novel method in which the {\underline{\em brown dwarf luminosity function}} is used was recently proposed by \cite[Caiazzo \etal\ (2017)]{icea17}; their method may become feasible in the {\em James Webb Space Telescope} era.  

Perhaps the most traditional flavor of the ``vertical method'' uses the unevolved cluster MS as reference locus to ``anchor'' the CMD~-- an approach dating back at least to \cite[Sandage (1970)]{as70}, and which is commonly referred to as {\em MS fitting}. Again, local subdwarfs with measured parallaxes and known chemical composition are used as calibrators. {\cite[Sandage (1986)]{as86} actually classifies this as a ``horizontal'' method, because of the significant slope of the MS in a CMD (unlike the HB, which is basically ``flat'' in the visual). This also implies that any metallicity differences between the cluster and field stars must be properly accounted for, due to their strong impact on the colors. Conversely, subdwarfs with known parallaxes and metallicities can be used to check the color-temperature transformations that are employed to place model isochrones in the observational plane (e.g., \cite[VandenBerg \etal\ 2010]{dvea10}, \cite[2016]{dvea16}; \cite[Chaboyer \etal\ 2017]{bcea17}). 

A clear example of how the MS fitting method works in practice is provided by \cite[Gratton \etal\ (1997)]{rgea97}. Using subdwarf parallaxes obtained with the Hipparcos satellite, a high-resolution spectroscopic analysis of their chemical compositions, and a Monte Carlo approach to estimate the modeling errors, these authors find that the Milky Way's oldest, most metal-poor GCs have an age of $11.8^{+2.1}_{-2.5}$~Gyr. This work was later revisited in \cite[Carretta \etal\ (2000)]{ecea00}, who found a similar age for the oldest clusters: $11.9 \pm 2.7$~Gyr.  

\cite[Mar{\'{\i}}n-Franch \etal\ (2009)]{mfea09} obtained relative ages for Milky Way globulars using data from the {\em HST}/Advanced Camera for Surveys (ACS) Treasury Program (\cite[Sarajedini \etal\ 2007]{asea07}). Cluster ages were mainly derived in a relative sense, with their MS ridgelines compared against one another, and only a few clusters~-- 47~Tuc, NGC~6397, and NGC~6752~-- used to pin down the absolute age scale, based on the MS fitting to local subdwarfs carried out for these three clusters by \cite[Gratton \etal\ (2003)]{rgea03}.  

\cite[O'Malley \etal\ (2017)]{omea17} obtained ages and distances of 22 GCs using subdwarfs with Gaia DR1 parallaxes as calibrators, and employing a Monte Carlo technique to assess the errors stemming from uncertainties in the input physics (see also \cite[Chaboyer \etal\ 1996a]{bcea96a}; \cite[Gratton \etal\ 1997]{rgea97}). They find ages in the range between $10.8$ and $13.6$~Gyr, with an average uncertainty of 1.6~Gyr. Similarly to \cite[VandenBerg \etal\ (2013)]{dvea13}, their results also support the existence of an age-metallicity anticorrelation, but do not clearly reveal the presence of a young outer-halo component. The latter result may be due to the relatively small sample sizes, compared with other studies in which such a component was clearly seen (e.g., \cite[Mar{\'{\i}}n-Franch \etal\ 2009]{mfea09}; \cite[Dotter et al. 2010,]{adea10} \cite[2011]{adea11}).

In the {\underline{\em ``horizontal'' methods}, it is the {\em color} of the MS TO, rather than its absolute magnitude, that is used as age tracer (\cite[Sarajedini \& Demarque 1990]{sd90}; \cite[VandenBerg \etal\ 1990]{vbs90}). Given the uncertainties that affect the temperatures of stellar models and the color-temperature transformations, such methods are generally considered suitable for the determination of relative ages only (see, e.g., \cite[VandenBerg \etal\ 1996]{vbs96}, \cite[2013]{dvea13}, where an extensive analysis of the impact of variations in different chemical abundances and physical ingredients is carried out).

{\em Isochrone fitting methods}, and more specifically those in which the morphology of the fit is used as a deciding criterion to infer the age, constitute a ``hybrid'' of sorts between vertical and horizontal methods, in the sense that they simultaneously use both magnitudes and colors as age diagnostics. As already pointed out by \cite[Iben \& Renzini (1984)]{ir84}, however, such methods must be used with caution, as some properties of stellar models, particularly their temperatures (hence colors), are subject to considerable uncertainties, stemming for instance from the treatment of convection in the modeling. There is also considerable debate as to the most appropriate transformations between model temperatures and colors (e.g., \cite[VandenBerg \etal\ 2016]{dvea16}; \cite[Chaboyer \etal\ 2017]{bcea17}; VandenBerg \& Denissenkov 2017, in prep.). Examples of ages recently derived using isochrone fitting include the studies by \cite[Dotter \etal\ (2010,]{adea10} \cite[2011)]{adea11}, which were based primarily on data collected for several dozen GCs with the ACS camera onboard {\em HST}. These studies confirm that the halo contains a dominant old component with ages of order 13~Gyr at the high end, but with a number of globulars that may be several Gyr younger, particularly in the outer halo. Interestingly, several of these studies suggest that the age-metallicity relation is actually quite flat for metallicities ${\rm [Fe/H] \lesssim -1.8}$~dex, a noticeable slope developing only at higher metallicities (see also \cite[VandenBerg \etal\ 2013]{dvea13}). 

In an interesting recent study, \cite[Wagner-Kaiser \etal\ (2017)]{wkea17} proposed a Bayesian fit of isochrone model grids to the observed CMDs of GCs as a method to simultaneously infer their ages {\em and} helium abundances $Y$. Applying the method to a sample of 69 globulars with ACS photometry, they obtain some intriguing results, including a mean $Y > 0.30$ (with individual values typically having errors $< 0.01$ in $Y$) and ages with reported errors that are often better than 0.1~Gyr. For M92 (NGC~6341), for instance, they find $Y = 0.347 \pm 0.004$ and ${\rm age} = 13.498^{+0.002}_{-0.007}$~Gyr. These results more likely reflect the dangers involved in this kind of approach, given the considerable uncertainties still affecting model isochrones, than a genuine triumph in recovering ultra-accurate parameters of star clusters. In particular, such high $Y$ values do not seem compatible with other He-sensitive features in the CMDs, such as the relative proportions between stars in the red giant branch and the HB (``$R$ method''; \cite[Iben 1968]{ii68}), the luminosity of the HB, and the periods of RR Lyrae stars. In fact, such diagnostics have generally not lent support to the existence of large cluster-to-cluster variations in $Y$, nor to metal-poor GCs having median $Y$ values significantly increased in comparison with the $Y \approx 0.24-0.25$ provided by Big Bang nucleosynthesis (BBN; e.g., \cite[Sandquist 2000]{es00}; \cite[Salaris \etal\ 2004]{msea04}; \cite[Gratton \etal\ 2010]{rgea10}; \cite[VandenBerg \etal\ 2016]{davb16}; but see \cite[Denissenkov \etal\ 2017]{pdea17}, in the case of M13 [NGC~6205]). This, of course, does not preclude the existence of subpopulations with significantly enhanced $Y$ in some globulars; even in those cases, however, stars with $Y > 0.30$ seem to be the exception, rather than the norm (\cite[Gratton \etal\ 2010]{rgea10}). 

\cite[Wagner-Kaiser \etal\ (2017)]{wkea17} also report a large number of GCs with ages that are very close to 13.49~Gyr, all with very small reported error bars. This result may be a consequence of their assumption of a ``hard'' upper limit to cluster ages, as provided by analyses of the cosmic microwave background. Indeed, different fits to $\Lambda$-Cold Dark Matter ($\Lambda$CDM) models combining {\em Planck} data with other relevant datasets (e.g., from weak lensing) all provide an age for the Universe of about $t_0 = 13.80 \pm 0.03$~Gyr (\cite[Planck Collaboration 2016b]{planck16b}). The latter is indeed an impressive result that provides a tight constraint, but unfortunately, apart from the fact that our stellar models are still admittedly imperfect (see, e.g., \cite[Valle \etal\ 2013]{gvea13}, \cite[2015]{gvea15}; \cite[Cassisi \etal\ 2016]{scea16}),\footnote{A reminder of the imperfection of our models is their current difficulty in reproducing the seismological data for our Sun~-- the so-called ``solar abundance problem'' (for reviews, see, e.g., \cite[Guzik 2008]{jg08}; \cite[Catelan 2013]{mc13}; \cite[Basu 2016]{sb16}; \cite[Serenelli 2016]{as16}). The latest (but still unsuccessful) efforts towards solving the problem, using for instance updated opacities, can be found in \cite[Le Pennec \etal\ (2015)]{mlpea15}, \cite[Buldgen \etal\ (2017)]{gbea17}, and \cite[Vinyoles \etal\ (2017)]{nvea17}, among others. The problem propagates to the low-metallicity regime also in the sense that abundances are typically given with reference to the solar value, and the latter may assume different values for different authors.}  there is no practical way our stellar evolution models can be ``informed'' {\em a priori} that stellar ages cannot be allowed to exceed a certain value. The controversy regarding possible variations in the mixing-length parameter of convective theory as a function of metallicity and position in the H-R diagram (e.g., \cite[Bonaca \etal\ 2012]{abea12}; \cite[VandenBerg \etal\ 2013]{dvea13}; \cite[Trampedach \etal\ 2014]{rtea14}; \cite[Magic \etal\ 2015]{zmea15}; \cite[Tayar \etal\ 2017]{jtea17}) further illustrates the difficulties that still affect the comparison between model predictions and the empirical data, particularly when color/temperature information is involved.\footnote{Note, in this sense, that different authors often adopt different $T\!-\!\tau$ relations, and even the gray approximation is still in use. By affecting the structure of the outer layers of stars, this also affects the calibration of the mixing-length parameter, with consequences upon the predicted $T_{\rm eff}$ values that propagate well beyond the solar case.} 

A reasonable approach, then, is to allow ages to fluctuate as one's specific isochrone set may favor (depending on the input physics and assumptions adopted in the modeling, in addition to the color-temperature transformations used to place the models in the observational plane). One may then choose to normalize the thus derived ages for better (nominal) agreement with the {\em Planck} results, and/or work with age ratios, as done for instance by \cite[Mar{\'{\i}}n-Franch \etal\ (2009)]{mfea09} and \cite[VandenBerg \etal\ (2013)]{dvea13}. I will come back to the subject of $\Lambda$CDM constraints in \S\ref{sec:lcdm} below.  

A potentially very powerful means of deriving accurate ages of GCs is by using detached, double-lined {\underline{\em eclipsing binary systems}} in which at least one of the stars is currently at the TO point, or in its immediate vicinity (\cite[Paczy\'{n}sky 1997]{bp97}). Nature has been very generous in providing us with a number of known such systems, such as those in the GCs $\omega$~Cen (NGC~5139; \cite[Thompson \etal\ 2001]{itea01}) and 47~Tuc (\cite[Thompson \etal\ 2010]{itea10}). The ages of these globulars can be inferred from the parameters (masses, radii,  luminosities) that can be derived from fits to both the photometry and spectroscopy. The possibility of obtaining ages directly from the masses is particularly attractive, as the latter are basically immune from the usual uncertainties affecting color-temperature transformations, reddening corrections, and the treatment of convection in the outer layers of low-mass stars. In this sense, detached eclipsing binaries also offer a unique opportunity to test our evolutionary models of low-mass stars. Naturally, the same approach can also be used in the case of higher-mass stars in younger systems (see, e.g., \cite[Brogaard \etal\ 2012]{kbea12}, in the case of the old, metal-rich open cluster NGC~6791). 

Using a Monte Carlo approach to evaluate the impact of uncertainties in the input physics, and also taking into account the observational errors, \cite[Chaboyer \& Krauss (2002)]{ck02} found an age of $11.10 \pm 0.67$~Gyr for the eclipsing system OGLEGC\,17 in $\omega$~Cen. Analysis of the eclipsing system V69 in 47~Tuc, on the other hand, suggests that some caution would be appropriate, even when both components of a system are very close to the MS TO, and their masses are known to within about $\pm 0.005 \, M_{\odot}$. Indeed, \cite[Thompson \etal]{itea10} derived ages for the V69 system using five different sets of isochrones, and based on different diagnostics~-- TO mass only, the mass-radius relation, and the mass-luminosity relation. In each case, the spread in inferred ages (reaching up to 2.8~Gyr, depending on which of the three diagnostics is used) is surprisingly large. The results also depend significantly on {\em which} of the two stars is used in the analysis, when only the stellar mass is used: differences at the level of 0.8 to 1~Gyr are found, depending on which models are adopted. According to \cite[Thompson \etal]{itea10}, the differences in inferred ages, from one model to the next, can be mostly traced to the different helium abundances that are adopted in the different models, even though the range in $Y$ among the five sets that they used is only $\Delta Y = 0.012$. Such a strong dependence on $Y$ is not welcome news, as far as the accuracy of this age-dating method is concerned, since $Y$ is a parameter that cannot be directly measured for such moderately cool stars. To alleviate this complication, \cite[Brogaard \etal\ (2017)]{kbea17} recently used the CMD morphology of the cluster around the TO-subgiant region to help further constrain the possible solutions. In their analysis, they also explored the impact of variations in [Fe/H], [$\alpha$/Fe], and [O/Fe]. Their best estimate of the cluster age is 11.8~Gyr. They also find that the cluster is likely older than 10.4~Gyr, at the $3\sigma$ level. However, the authors caution that, depending on the exact adopted $T_{\rm eff}$ scale and He abundance, the favored age can rise up to the 12.5 or 13.0~Gyr mark.

The detailed chemical composition, then, including $\alpha$-to-iron ratios and He abundances, can affect the ages that one infers from analyses of the properties of stars around the TO and subgiant regions. In this sense, it is worth recalling that there are widespread (and systematic) differences in $[\alpha/{\rm Fe}]$ at fixed [Fe/H] between stars in dwarf galaxies (including the Magellanic Clouds and their GCs) and those in the Milky Way halo (e.g., \cite[Venn \etal\ 2004]{kvea04}; \cite[Tolstoy \etal\ 2009]{etea09}; A. Mucciarelli 2017, priv. comm.).  At higher metallicities, in turn, both toy models (\cite[Catelan \& de Freitas Pacheco 1996]{cdfp96}; \cite[Catelan 2009b]{mc09b}) and realistic simulations (A. Wetzel 2017, priv. comm.) predict the existence of different stellar populations with different $[\alpha/{\rm Fe}]$ {\em and} different helium abundances $Y$, again at fixed [Fe/H] (see also \cite[Nataf 2016]{dn16}, and references therein). In order to derive their ages, {\em even in a relative sense}, such chemical composition variations must be carefully taken into account, lest systematic errors that can potentially exceed 1~Gyr (depending on the method) be incurred. In this context, it would also be important to settle the debate regarding the degree to which [O/Fe] may increase at ${\rm [Fe/H]} \lesssim -2$, given its known impact on stellar ages (\cite[VandenBerg \etal\ 2016]{dvea16}, and references therein).

\section{The ages of the oldest stars and the $\Lambda$CDM paradigm}\label{sec:lcdm} 

Halo subgiants with well-measured parallaxes and known chemical composition are also good candidates to provide an accurate lower limit to the age of the Universe. \cite[VandenBerg \etal\ (2014)]{dvea14} analyzed three such stars for which parallaxes could be obtained from extensive observations carried out with the Fine Guidance Sensors onboard {\em HST}. For one of these subgiants, HD\,140283, they favored an age of $14.27 \pm 0.38$~Gyr, which is nominally $0.47 \pm 0.38$~Gyr older than the age of the Universe favored by the {\em Planck} results (\cite[Planck Collaboration 2016b]{planck16b}; see also \cite[Bennett \etal\ 2013]{cbea13}, in the case of {\em WMAP}). However, the epoch of reionization is also constrained to have begun at $z = 8.9^{+3.3}_{-1.8}$ (\cite[Planck Collaboration 2016a]{planck16a}); this implies that the first stars only started forming in earnest some $0.56 \pm 0.20$~Gyr after the Big Bang, at an age $t_{0,\star} = 13.24 \pm 0.20$~Gyr. This differs from the estimated age of HD\,140283 by $1.03 \pm 0.43$~Gyr, a $1.4 \, \sigma$ result. 

Interestingly, the Gaia DR1 parallax for this star is inconsistent with the {\em HST} value, which had been thought to be particularly accurate. On the other hand, as shown in Figure~\ref{fig:fig1}, the {\em HST} parallax is fully consistent with the Hipparcos value for the same star. If correct, the Gaia result would imply a significant reduction in the star's age, essentially removing any discrepancy with the {\em Planck} result. Given the known caveats associated with Gaia DR1 (\cite[see Eyer \etal\ 2017]{leea17}), its next data releases are eagerly anticipated. If the latter happen to confirm the previous {\em HST} and Hipparcos value, however, further work will be required to properly establish the reasons for the star's high inferred age. There is significant potential for stars such as HD\,140283 to teach us about the limitations of our current stellar models. With some luck (though much less likely), we may even be able to learn something new about cosmology in the process, as even the remarkable results from {\em WMAP} and {\em Planck} are based on their own sets of assumptions (\cite[Soderblom 2010]{ds10}; \cite[Valls-Gabaud 2014]{dvg14}). Note, in this sense, that \cite[Creevey \etal\ (2015)]{ocea15} obtained an age for HD\,140283 between 12.2 and 13.7~Gyr, depending on the assumed extinction, with errors $\sim \pm 0.6-0.7$~Gyr. However, their result was challenged by \cite[VandenBerg \etal\ (2016)]{dvea16}, on the basis of the adopted thermonuclear reaction rates, low $T_{\rm eff}$ (which implies an unrealistic value of the mixing length parameter), and [O/Fe] ratio. Still, \cite[VandenBerg \etal\ (2016)]{dvea16} recently pointed out that a more realistic treatment of the equation of state would already lead to a reduction of the age of HD\,140283 down to about 14.0~Gyr.

\begin{figure}
\floatbox[{\capbeside\thisfloatsetup{capbesideposition={right,top},capbesidewidth=4.75cm}}]{figure}[\FBwidth]
{\caption{Parallax measurements for HD\,140283. The Gaia value (TGAS in the plot, after Tycho-Gaia Astrometric Solution) deviates significantly from the one obtained with {\em HST} in particular, which in turn is fully consistent with the Hipparcos value.}\label{fig:fig1}}
{\includegraphics[width=7.5cm]{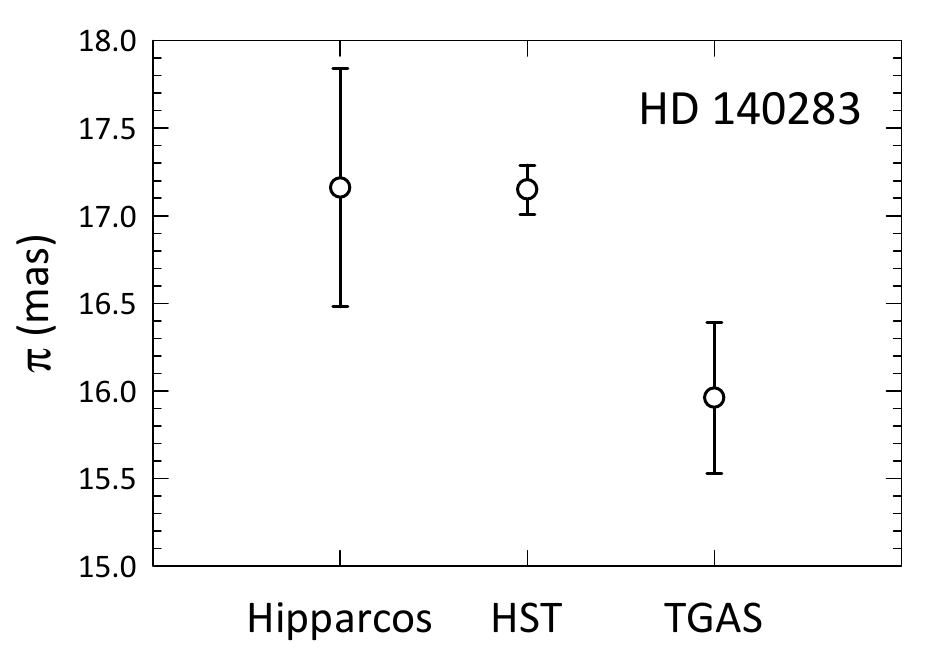}}
\end{figure}

\section{Nucleocosmochronology and the ages of the oldest stars}\label{sec:nuc}

An independent route towards the ages of the oldest halo stars is provided by nucleocosmochronology (for recent reviews, see \cite[Frebel \& Katz 2009]{fk09}; \cite[Christlieb 2016]{nc16}). The technique works not unlike what was described in \S\ref{sec:sun} for our Sun, only here, instead of measuring the abundances of the different isotopes of (say) U, Th, and Pb in meteorites, one must measure the {\em global} abundances of these elements directly from high-resolution stellar spectra. This is a difficult task, considering that the stars in question are very metal-poor, and the very few relevant lines that can be used are typically weak and may also be blended with other lines. For instance, the U\,{\sc ii} line at 3859.57\,\AA, which is the only U line in the visible that has been used so far in this context, is typically blended with C in C-enhanced stars. When U is not available, but Th is, one often resorts to comparing the latter to other $r$-process elements, such as Eu or Os. Either way, the production ratios must be known a priori\,--\,a challenging task, considering that there is not even a consensus on what the actual source(s) of $r$-process enrichment may be. It does not come as a surprise, therefore, that the uncertainties in these key production ratios remain very high, but they are even more uncertain when comparing nuclides with large mass differences (\cite[Christlieb 2016]{nc16}). Thus, the Th/U chronometer is thought to constitute the more dependable one at present (\cite[Sneden \etal\ 2003]{csea03}; \cite[Frebel \& Katz 2009]{fk09}, and references therein). 

\cite[Christlieb (2016)]{nc16} summarizes some of the most recent age measurements obtained using nucleocosmochronology, for six halo stars with $-3.0 < {\rm [Fe/H]} < -2.0$. The individual values range from 11.7 to 14.8~Gyr, with error bars ranging between 2 and 4~Gyr. A weighted mean over the compiled values formally gives an age of $13.5 \pm 1.0$~Gyr, but the actual error bar is dominated by systematics, and likely exceeds several Gyr.

\section{Concluding remarks}\label{sec:conc}

The current evidence, from the age-dating of field halo stars and GCs alike, strongly suggests that the oldest Milky Way stars have ages that are comparable to the age of the Universe. All available techniques, however, still suffer from a variety of error sources, including systematic ones, that render absolute ages uncertain, probably at the $\gtrsim 2$~Gyr level (depending on the technique). Much work will still be necessary, before stellar ages can achieve an accuracy level that will make them useful in constraining cosmological models, {\em vis-\`a-vis} the results from the {\em Planck} mission.

\acknowledgments I warmly thank Don A. VandenBerg for many interesting discussions, and for a critical reading of an early draft of this paper. I also thank him, A. Mucciarelli, and A. Wetzel for sharing some of their results in advance of publication. The following agencies and grants are also acknowledged: Millennium Science Initiative grant IC\,120009, awarded to the Millennium Institute of Astrophysics (MAS); Proyecto Basal PFB-06/2007; FONDECYT grant \#1171273; and CONICYT's PCI program through grant DPI20140066. 


\end{document}